# Preprint

# A Knowledge Graph-Based Method for Automating Systematic Literature Reviews

Nada Sahlab*, Hesham Kahoul,* Nasser Jazdi*, Michael Weyrich*

*University of Stuttgart, Institute of Industrial Automation and Software Engineering
Pfaffenwaldring 47, 70569 Stuttgart, Germany. Firstname.lastname@ias.uni-stuttgart.de

**Abstract**

Systematic Literature Reviews aim at investigating current approaches to conclude a research gap or determine a futuristic approach. They represent a significant part of a research activity, from which new concepts stem. However, with the massive availability of publications at a rapid growing rate, especially digitally, it becomes challenging to efficiently screen and assess relevant publications. Another challenge is the continuous assessment of related work over a long period of time and the consequent need for a continuous update, which can be a time-consuming task. Knowledge graphs model entities in a connected manner and enable new insights using different reasoning and analysis methods.

The objective of this work is to present an approach to partially automate the conduction of a Systematic Literature Review as well as classify and visualize the results as a knowledge graph. The designed software prototype was used for the conduction of a review on context-awareness in automation systems with considerably accurate results compared to a manual conduction.

## 1. Introduction

Literature reviews are a significant part of a research activity, based on which an overview on a certain research topic can be given as well as conclusions drawn for the assessment of own work. Some methods exist for conducting a systematic literature review, where the process is divided into several steps starting with the design, the conduction of the review as well as the analysis and documentation of results. Some of the notable methods are presented in [1] and [2]. So far, systematic literature reviews are conducted manually, which can be an effortful task, especially due to the wide search scope across digital libraries and the necessary filtration of findings. With regard to electronic databases containing publications, a continuous growth of stored publications makes a correct assessment, filtration and selection of related work cumbersome, even with using available extended search strings. As research is evolving and continuously being extended with new approaches and work, an automated support for conducting a literature review is needed. Conducting a systematic literature review is usually an activity for early or late research phases. Having a continuously updated view on the progression of research with regard to a certain topic can be beneficial in benchmarking the own approach and finding correlations or discrepancies. Therefore, a need to represent publications in a connected manner, while relating them to specific aspects within a research topic arises. Several approaches such as connected papers [3] and Academic Knowledge Graph [4] present publications as a connected graph based on topic similarity in order to give researchers an overview on the relatedness of subjects and authors as well as highlight on influential and highly cited works. Mostly, the research topics presented by these approaches are general and not tailored to a specific literature review. Knowledge graphs (KG) define a network of domain information represented as nodes and connected via edges, which evolve and allow inferences. Using a knowledge graph to represent findings for a systematic literature review enables efficient analysis as well as improved conclusions.

The objective of this work is therefore twofold: First to present an approach to partly automate the conduction of a systematic literature review by designing interfaces to digital libraries and embedding search criteria as well as using natural language processing to further filter and classify results. The second objective is to present publications as a knowledge graph and enable an enhanced analysis by querying the graph. The rest of the paper is structured as follows, in section 2, an overview on systematic literature reviews as well as related work is provided. In section 3, the concept is described and presented with a use case in section 4. An evaluation of the methodology is provided with regard to the accuracy of the results in section 5. To conclude, section 6 provides a summary and an outlook.

## 2. Preliminaries

*2.1. Systematic Literature Reviews*

According to [5], a Systematic Literature Review (SLR) is a strategy to recognize, assess and summarize the state-of-the-art of a particular subject within the literature. SLRs collect publications, literature and information, allowing a thorough methodological and objective investigation as opposed to conventional surveys. An SLR has the goal of forming a ground for a specific topic and work out a systematic outline for works related to it in literature.

As mentioned in [1, 6], an SLR consists of three consecutive parts: planning the review, conducting the review and documenting it (Fig 1.). In the first phase, a research question is formulated and further steps for carrying out the review are designed and iteratively validated. Depending on the defined research questions, the following steps are conducted. The conduction phase follows, where the literature review takes place. To constrain the search scope, inclusion and exclusion criteria are defined, which limit the review to certain types of literature, a certain domain or a time-limit for the review. For the search, a backward or forward search is possible. Backward searches rely on cited work in specific publications whereas forward search is by using search engines for instance.

Keywords used in the search process play a vital role in the identification of correct literature. The keywords should be extracted from the research questions which are framed previously. Going through the different publications and acquiring the data is conducted in 2 steps, where first an initial screening of the abstract and if relevant a detailed reading of the publication takes place. Quality assessment criteria are defined to have an evaluating schema for the acquired data and its analysis. For the synthesis of data, statistical approaches can be applied to show the results as well as further descriptive methods.

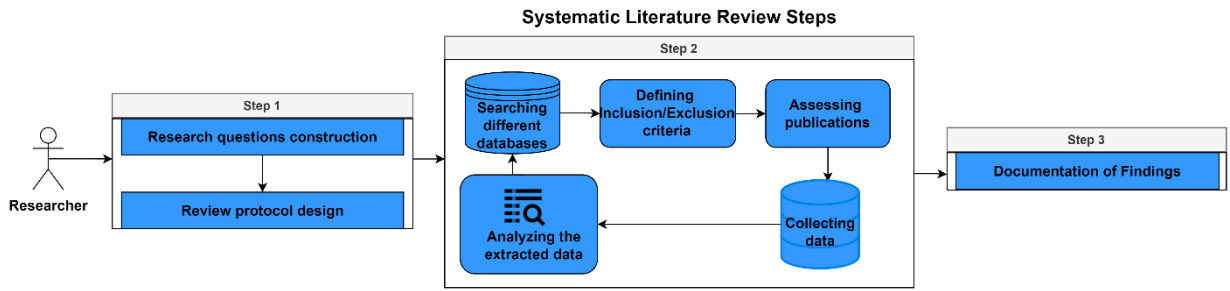

Figure 1 Systematic Literature Review Stages

## 2.2. Academic Knowledge Graphs

In recent years, some approaches to visualize and create bibliometric networks were presented. The aim of these approaches is to visualize papers, while highlighting on their connectedness and similarity. In the following sections, some approaches are discussed.

### 2.2.1. Connected Papers

Connected Papers is a connected graph approach to show similarity between publications even if there is no direct citation and to infer topic similarity. Publications which do not explicitly cite one another can also be closely linked and placed together in the graph. Technically, the resemblance of the paper in the graph is mostly centered on the notions of bibliographic linkage and co-citation. As per this criterion, two articles with a lot of identical citations and references are more likely to deal with the same topic area.

The Connected Papers graph is developed in such a way that it draws attention to the most relevant and essential papers straight away. The graph has an algorithm in the form of a layout, where identical papers are clustered together in space and are further connected together by thick lines called edges. Frequently cited papers are depicted by bigger nodes and the recently published papers are indicated in a darker color. [3]

### 2.2.2. AceMaps

Another approach to build a knowledge graph based on literature publications is the open-source Multimodal Geoscience Academic Knowledge Graph [7], which aims at encompassing domain knowledge in different research fields.

The approach offers an ontology with millions of entities. The graph entities include information about the affiliation, author and so on. The ontology and the dataset are open to use, and this approach provides insights about trending fields and authors based on the modeled knowledge about their publications and citations.

### 2.2.3. VosViewer

VosViewer [8] is a tool for mapping and constructing bibliometric networks based on co-authorship or co-occurence. The approach focuses on distance-based visualization of literature to indicate similarity between publications, using which, publications in a specific field with high impact can be identified.

### 2.2.4. Other Related Work

In [9], an approach for automating the structuring and clustering of concepts related to specific research topics is presented. SciKGraph is a software that aims at automating the structuring of concepts and topics within a scientific domain using a semantic-based analysis and clustering mechanisms. Documents are classified using Natural Language Processing with results showing high accuracy.

In [10], an ontology for semantic surveys (SemSur) is presented with the aim of altering the way researchers conduct and present surveys by using a common base and presenting findings as a knowledge graph. Using the content of some survey papers, useful information could be extracted with the help of queries. A contribution was therefore made to how survey papers are presented.

*2.2.5. Concluding Remarks*

The presented approaches show how a presentation of connectedness and similarity among publications can be of support for researchers to find influential publications and expand their base of related work. Other works highlight on the common understanding using semantic technologies. Although providing support and proposing paradigm shifts, the approaches are mainly visual or consider the extraction of information from already present publications. Extending this support by customizing and presenting search results from various digital libraries for a systematic literature review is a further step addressed by the proposed concept, which will be described in the following chapter.

## 3. A Knowledge-Graph Based Methodology for Automating Systematic Literature Reviews

*3.1. Concept Overview*

In the following section, the approach of partially automating the SLR conduction is presented.

As mentioned in [1, 6, 11], it is essential as a first step of SLR, to formulate research questions (RQ). As shown in Fig.1, the first stage of the SLR cannot be automated as researchers need to identify what they want to capture from the SLR process and why. The second stage of the SLR consists of five different sub-stages. The first sub-stage is identifying the channels for literature search. As mentioned in [5], there are three main ways to search for data concerning certain topics, which can be via backward searching, forward searching and electronic databases. In this study, searching the electronic databases will be used to automate the process of the SLR.

Consequently, the inclusion/exclusion criteria must be identified. All electronic databases that are used in this study provide advanced search that helps in including only certain publications that are studying some topics. In order to use the advanced search, Boolean operators are used between search terms such as (AND & OR) operators to only search for specific keywords that are extracted from the research questions. Stage 2 is automated with regard to the collection of data and its representation as a knowledge graph. The quality assessment, which requires reviewing the full text manually by the researcher is therefore not considered. For synthetizing and presenting the results, a partial automation via generic queries of the knowledge graph is possible.

*3.2. Data Acquisition*

In order to conduct SLR automatically, a python-based interface is used to extract data from different electronic databases for applying further analysis on them. Here, different keywords and the inclusion/exclusion criteria are embedded them into the database URL. The HTML of the database are then parsed, and specific data extracted. Table 1 represents the data extracted from the publications within the databases.

| No. | Extracted data |
|---|---|
| 1 | Publication title |
| 2 | Author names |
| 3 | Abstract |
| 4 | Publication year |
| 5 | Type of the publication |
| 6 | Keywords |
| 7 | Citation |

Table 1: Data items extracted from the literature

After collecting the data from the publications, natural language processing (NLP) techniques are used on the title, the abstract and the keywords to reach some knowledge from the extracted data. NLP methods should be able to extract the field of application of the literature and some entities or keywords related to the publication. The results are then stored in CSVs which are then used to build a KG based on a pre-defined meta-model.

Using NLP, data is arranged and unnecessary text that is not important in the analysis such as subject pronouns, prepositions, math operators and equations and other special characters like "§", "%", "/", etc. is removed. After this process, a clean text which is ready to be analyzed is obtained. The next step is the classification of the publication and extraction of entities based on pre-trained deep learning models.

*3.3. Knowledge Graph Metamodeling*

The knowledge graph meta-model is designed based on schema-based design. In order to use schema-based design, a specific model should be designed with no ambiguities.

Fig. 2 shows the visualization of the expected knowledge graph with appropriate nodes and edges. Here, a labelled property graph (LPG) is used, consisting of labelled nodes and edges. LPGs have the advantages of assigning key-value properties to both nodes and edges, which can be helpful later on in analysing the graph by weighing edges and further categorizing nodes by labels.

As shown, the data extracted from the publications are stored in nodes and relationships are created to connect each node with each other. Each node stores certain information as discussed below:

- Publication node: this node stores information related to the publication such as:
    - Name of the authors
    - The abstract
    - The title
    - The URL
    - Keywords
- Year of publication node: the year that the publication published in is stored in this node.
- Database node: this node stores the database that publishes the publication. The label of this node should be one of the five databases mentioned before.
- Citation node: the number of citations that the literature was cited is stored in this node
- Related keywords node: it stores the entities that are related to the publication.
- Field of application node: the scope of the publication is stored in this node.

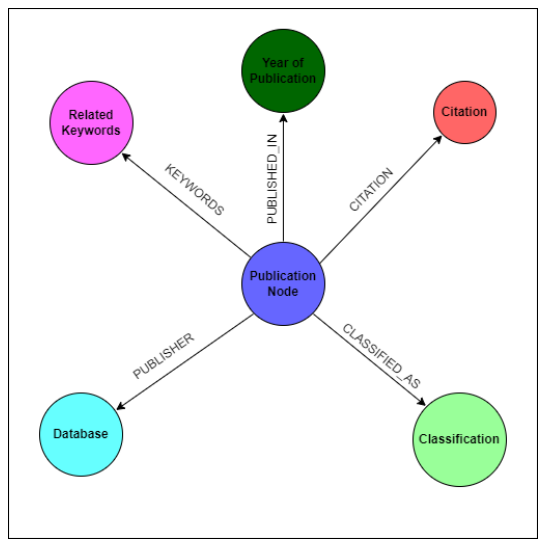

Figure 2: KG meta-model

*3.4. Analysis*

*3.4.1. Designing the queries to analyze the extracted publications*

After building the KG, queries should be implemented to gain insights about each publication as well as analyze publications from different views.
Here, two different types of generic queries were designed.
The pre-designed queries allow the user to cluster the KG by:
- The field of application or scope each publication considers
- The year that the literature is published in
- The database where the literature is published

Furthermore, the application also allows the user to design his/her own queries to further analyze the KG. For example, sorting publications based on the number of citations or searching for specific authors.

*3.4.2. Knowledge Graph Update*

Knowledge Graphs have numerous advantages as they are dynamic and have the ability to change according to the change in the data. They are also expandable and updatable which make them very efficient in showing new entries and drawing relationships between the new and the old data that are already existing [12, 13].

This work also considers the importance of expandability of the KG. Whenever the user triggers new search, the already existing KG will be updated and expanded automatically which in turn makes it very efficient.

## 4. Realization

*4.1. Workflow*

In this section, the components of the prototype to automate a systematic literature review using knowledge graphs is discussed. As shown in Fig.3, the first step is to collect publications from different databases such as IEEExplore, Springer link, Wiley, ACM and ScienceDirect. In order to collect the publication from databases and extract useful data from them, python libraries are used to automate this process. Then, the extracted data are stored in CSVs to help building a knowledge graph in the graph-database neo4j. Using the graph query cypher language, graph queries were designed.

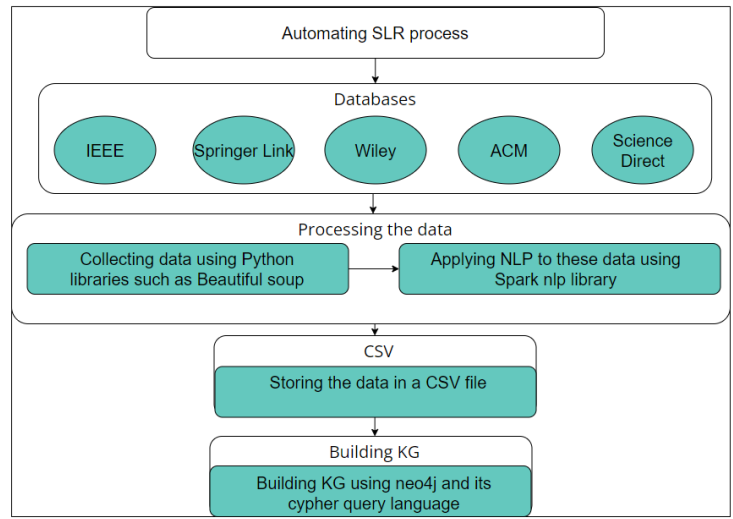

Figure 3: Workflow

*4.1.2 Front-end design of the Web-Application*

A Web-Application was realized to facilitate the search and visualization of the SLR. The Web-Application takes in the search keywords and presents the user with the knowledge graph, which is saved in the graph-database. The Web-Application consists of text boxes to specify the search terms as well as the exclusion criteria such as specifying the range of the years and/or using logic operators to include certain publications. Fig. 4 shows a screenshot of the front-end.

Figure 4 Front-end web application

*4.2. Use Case*

As a proof of concept, the Web-Application is used to conduct an SLR in the field of context-awareness in automation systems. The research questions for the SLR aimed at investigating context modeling and reasoning approaches as well as classify application fields within the automation technology, where context-aware systems were applied. The review considered publications of the past five years. Fig.5 shows the front-end when the user specifies his search terms to be "context-awareness AND automation systems" and the year range to be from 2016 to 2022.

Figure 5: search keywords for the use case

*4.2.1. Capturing the results of the SLR using KG automatically*

Fig.6 shows how one publication node and its relationship with other nodes should look like. While Fig.7 shows a part of the KG.

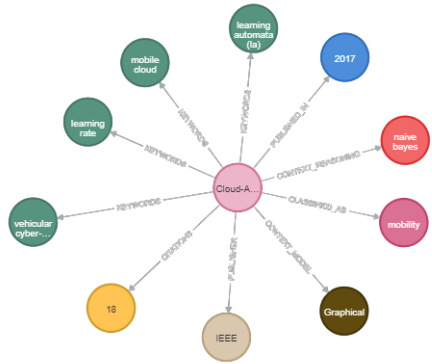 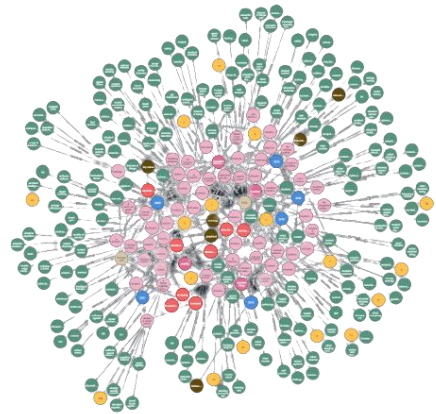

Figure 6: One publication          Figure 7: Part of the KG

*4.2.2. Evaluation*

A manual SLR was conducted on context awareness in automation systems prior to using the realized prototype. The results of the manual SLR were then used as a benchmark to test the accuracy of the automated SLR.

The comparison of the manual and automatic search shows a high accuracy and filtering of unsuitable publications. Automatic SLR managed to extract 78% of the publications that are resulting from manual SLR. The accuracy of the classified publications using automatic SLR is about 90% similar to manual SLR. Moreover, both manual and automatic SLR enable the same conclusion about prominent context modelling and reasoning approaches that are used within the publications in the past 5 years and in which field of application within automation systems.

Overall, a total of 212 publications were acquired and filtered from a total of 36973 publications found via the search engines of the five databases. A manual filtration yielded a total of 153 publications considered. The manual SLR, which included quality assessment criteria, considered 52 publications in total. Table 2 sows a summary of the extracted data using automated SLR.

| No. | Database | After Search | After Inclusion/Exclusion | Accuracy of the automated SLR with respect to manual SLR (benchmark) | |
|---|---|---|---|---|---|
| | | | | Extracted publications similarity | Classifying the publications using NLP |
| DB1 | IEEEXplore | 6798 | 45 | 57% | 89% |
| DB2 | Springer Link | 9634 | 53 | 92% | 91% |
| DB3 | Science Direct | 8825 | 29 | 64% | 100% |
| DB4 | ACM | 9726 | 17 | 90% | 78% |
| DB5 | Wiley | 1990 | 9 | 100% | 100% |
| | TOTAL | | 153 | 78% | 90% |

Table 2 Summary of the extracted data using automated SLR

## 5. Summary and Conclusions

In this contribution, an approach for automating the data acquisition step of a Systematic Literature Review and presenting the results as a connected knowledge graph was presented. The proposed approach automates the initial steps for searching for relevant publications, while considering inclusion and exclusion criteria specific to the individual research scope. Through querying mechanisms, synthesizing results showing relations and similarities are possible with the proposed queries and key-word extraction mechanism. The automatic assessment via APIs enables researchers to initiate an update on the existing literature review results and can accumulate and analyze the results based on a larger period of time without the effort needed for the classical approach. The designed knowledge graph can be used for peer-assessment and enables a multimodal view on the extracted results for making a general statement about the research topic. A detailed analysis of each publication can further enhance the knowledge graph with inferred knowledge, for example, by weighing the edges or applying distance-based, co-occurrence and similarity measures.

The prototype is being optimized and enhanced by testing it for further use cases to enable a generalized use across different research domains.

## 6. References

[1] B. Kitchenham, O. Pearl Brereton, D. Budgen, M. Turner, J. Bailey, and S. Linkman, "Systematic literature reviews in software engineering – A systematic literature review," *Information and Software Technology*, vol. 51, no. 1, pp. 7–15, 2009, doi: 10.1016/j.infsof.2008.09.009.

[2] M. J. Page *et al.,* "The PRISMA 2020 statement: an updated guideline for reporting systematic reviews," *BMJ (Clinical research ed.)*, vol. 372, n71, 2021, doi: 10.1136/bmj.n71.

[3] Connected Papers, *https://www.connectedpapers.com/* (accessed: Apr. 25 2022).

[4] C. Deng *et al.,* "GAKG: A Multimodal Geoscience Academic Knowledge Graph," in *Proceedings of the 30th ACM International Conference on Information & Knowledge Management*, Virtual Event Queensland Australia, 2021, pp. 4445–4454.

[5] Y. Xiao and M. Watson, "Guidance on Conducting a Systematic Literature Review," *Journal of Planning Education and Research*, vol. 39, no. 1, pp. 93–112, 2019, doi: 10.1177/0739456X17723971.

[6] P. Brereton, B. A. Kitchenham, D. Budgen, M. Turner, and M. Khalil, "Lessons from applying the systematic literature review process within the software engineering domain," *Journal of Systems and Software*, vol. 80, no. 4, pp. 571–583, 2007, doi: 10.1016/j.jss.2006.07.009.

[7] Acemap Inc., *GAKG: Multimodal Geoscience Academic Knowledge Graph.* [Online]. Available: https://gakg.acemap.info/#/ (accessed: Apr. 25 2022).

[8] N. J. van Eck and L. Waltman, "Visualizing Bibliometric Networks," in *Measuring Scholarly Impact: Methods and Practice*, Y. Ding, R. Rousseau, and D. Wolfram, Eds., Cham, s.l.: Springer International Publishing, 2014, pp. 285–320.

[9] Mauro Dalle Lucca Tosi, Julio Cesar dos Reis, SciKGraph: A knowledge graph approach to structure a scientific field, Journal of Informatics, Volume 15, Issue 1, 2021, 101109, ISSN 1751-1577,https://doi.org/10.1016/j.joi.2020.101109.

[10] Fathalla, S., Vahdati, S., Auer, S., Lange, C. (2017). Towards a Knowledge Graph Representing Research Findings by Semantifying Survey Articles. In: Kamps, J., Tsakonas, G., Manolopoulos, Y., Iliadis, L., Karydis, I. (eds) Research and Advanced Technology for Digital Libraries. TPDL 2017. Lecture Notes in Computer Science(), vol 10450. Springer, Cham. https://doi.org/10.1007/978-3-319-67008-9_25

[11] K. Mattick, J. Johnston, and A. de La Croix, "How to…write a good research question," *The clinical teacher*, vol. 15, no. 2, pp. 104–108, 2018, doi: 10.1111/tct.12776.

[12] *Industry-scale knowledge graphs: lessons and challenges*, 2019. [Online]. Available: https://dl.acm.org/doi/fullhtml/10.1145/3331166

[13] A. Hogan *et al., Knowledge Graphs*: ACM Comput. Surv. 54, 4, Article 71 (May 2022), 37 pages. https://doi.org/10.1145/3447772